\documentclass[twocolumn,showpacs,preprintnumbers,amsmath,amssymb,floatfix]{revtex4}
\usepackage{graphicx}% Include figure files
\usepackage{dcolumn}% Align table columns on decimal point
\usepackage{bm}% bold math
\usepackage{fancybox}
\newcommand{\GWA}{{\it GW}A}

\newcommand{\bfq}{{\bf q}}

\newcommand{\bfr}{{\bf r}}

\def\qsgw{QS{\em GW}}

\def\ei{\varepsilon_i}

\def\ej{\varepsilon_j}

\def\H0{H^0}

\def\veff{V^{\rm eff}}
\def\vxc{V^{\rm xc}}

\newcommand{\req}[1]{\mbox{Eq.~(\ref{#1})}}

\def\efermi{\mbox{$E_{\rm F}$}}

\def\connect#1{\leavevmode{\setbox1=\hbox{#1}\copy1%
\raise .2\ht1 \vbox{\moveleft \wd1\vbox{\hrule width \wd1 height .5pt depth 0pt}}%
}}

\def\ftn[#1]{\rlap{\footnotemark[#1]}}

\def\abmo{La$_{1-x}$A$_x$MnO$_3$}

\def\min{\uparrow}
\def\maj{\downarrow}

\def\ttg{$t_{\rm 2g}$}
\def\eg{$e_{\rm g}$}

\begin{document}
%\special{papersize=8.5 in, 11 in}
\title{Re-examination of half-metallic ferromagnetism 
for doped LaMnO$_3$\\ in quasiparticle self-consistent $GW$ method}
\author{Takao Kotani}
\affiliation{Faculty of engineering, Tottori university, Tottori 680-8552, Japan}
\altaffiliation{moved from Arizona State University, Tempe, USA}
\author{Hiori Kino}
\affiliation{National Institute for materials science, Sengen 1-2-1, Tsukuba, Ibaraki 305-0047, Japan.}
\date{\today}

\begin{abstract}
We apply the quasiparicle self-consistent $GW$ (\qsgw) method 
to a cubic virtual-crystal alloy La$_{1-x}$Ba$_x$MnO$_3$ %(LBMO)
as a theoretical representative for colossal magnetoresistive 
perovskite manganites. 
The \qsgw\ predicts it as a fully-polarized half-metallic ferromagnet
for a wide range of $x$ and lattice constant. 
Calculated density of states and 
dielectric functions are consistent with experiments. 
In contrast, the energies of calculated spin wave 
are very low in comparison with experiments.
This is affected neither by  
rhombohedral deformation nor the intrinsic deficiency in the \qsgw\ method.
Thus we ends up with a conjecture that phonons related to the
Jahn-Teller distortion should hybridize with spin waves more strongly 
than people thought until now.

\end{abstract}

\maketitle
%%%%%%%%%%%% Intro %%%%%%%%%%%%%%%%%%%%%%%%%%%%%%%%
\section{introduction}
The mixed-valent ferromagnetic perovskite \abmo,
where A is an alkaline-earth such as Ca, Sr or Ba, 
shows the colossal magnetresistence, e.g, 
see reviews by Tokura and Nagaosa \cite{tokura00}, 
and by Imada, Fujimori and Tokura \cite{imada98}.
As they explain, the colossal magnetresitance 
is related to the complexed interplay of spin, orbital and lattice
degree of freedoms. This is interesting not only from a view of
physics, but also of its potential applicabilities.
This interplay can be also related to the fundamentals of high-$T_{\rm c}$
superconductors and the multiferroic materials, 
which are now intensively being investigated \cite{sangwook07,ramesh07}.

In order to understand the interplay, kinds of theoretical 
works have been performed until now. They can be classified into two approaches; 
one is model approaches, and the other is the first-principle 
ones which are mainly based on the density functional theory
in the local density approximation (LDA) or in the generalized gradient approximation (GGA)
 \cite{sawada97,solovyev96,solovyev99,fang00,ravindran02}.
The first-principle approaches have an advantage that it can give energy bands 
(as quasiparticles) without any knowledge of experimental input.
Then kinds of properties are calculated based on the quasiparticles. 
However, it is well known that the density functional theory in the LDA (and GGA)
often fails to predict physical properties for compounds including transition metals. 
For example, Terakura, Oguchi, Williams and J. K{\"u}bler
\cite{terakura84a,terakura84} showed that the density functional
theory in the LDA is only qualitatively correct for MnO and NiO. 
Calculated band gaps and exchange splittings are too small,
resulting in poor agreement with optical and spin wave experiments 
\cite{Faleev04,kotani07a,solovyev98mno,kotani08sw}. 
This is little improved even in the GGA.

As a remedy, the LDA+$U$ method has been often used \cite{anisimov97}. 
However, it has the same shortcomings as model calculations: 
It can contain many parameters which are not determined within the theory itself, 
e.g, different $U$ for \ttg\ and \eg\ orbitals
\cite{solovyev96,sawada98} and $U$ for O($2p$) (oxygen
$2p$)\cite{korotin00}. Even though there are theoretical efforts in progress to evaluate
these $U$ parameters in first-principle methods
\cite{kotani00,miyake:085122}, we now usually have to determine these
parameters by hand so as to reproduce some experiments in practice.
Then we need to check whether calculations with the parameters
can explain other experiments or not.

Many researches are performed along this line.
Soloveyv et al. investigated LaTO$_3$(T=Ti-Cu) in the LDA+$U$,
where they tested possible ways of LDA+$U$ in comparison with experiments.
Then they concluded that LDA+$U$ gives little differences from the
results in the LDA in the case of LaMnO$_3$. It followed by their successful 
description of the spin-wave dispersions \cite{solovyev99} and 
phase diagrams \cite{fang00} in the LDA even for $x\ne 0$.
Ravindran et al. also showed detailed examination for LaMnO$_3$ with
full-potential calculations including spin-orbit coupling 
and full distortion of crystal structure \cite{ravindran02}, 
where they concluded that the density functional theory 
in the GGA worked well for LaMnO$_3$.
Thus, both of these groups reached to the same conclusion that
``We can treat \abmo\ accurately enough with the density functional
theory in the LDA or in the GGA, do not need to use LDA+$U$.''. 
It sounds very fortunate because we are not bothered with difficulties 
about how to determine parameters $U$ in the LDA+$U$. 
However, we must check this conclusion carefully.
For example, one of the reason why the GGA is accurate is based on
their observation that their calculated imaginary part of dielectric function 
$\epsilon_2(\omega)$\ in the GGA agrees well with an experiment ~\cite{ravindran02}. 
However, this is not simply acceptable if we recall other cases where 
peaks in the calculated $\epsilon_2(\omega)$ are deformed and pulled down to lower energies 
when we take into account excitonic effects.
Thus it is worth to re-examine the conclusion by some other
methods which are better than those dependent on the LDA or the GGA.

Here we re-examine the conclusion by the quasiparticle 
self-consistent $GW$ (\qsgw) method, which is originally developed 
by Faleev, van Schilfgaarde, and Kotani \cite{Faleev04,kotani07a}.
Its theoretical and methodological aspects, and how it works 
are detailed in \cite{kotani07a} and references there in. 
They showed that the \qsgw\ method gave reasonable results 
for wide-range of materials. 

In Sec.\ref{method}, we explain our method. 
Then we give results and discussions in Sec.\ref{results}.
In our analysis in comparison with experiments, 
calculated quasiparticle energies given by the \qsgw\
seems to be consistent with experiments. However, the obtained spin
wave energies are about four times too larger than experimental values.
From these fact, as for \abmo, we ends up with a conjecture that phonons
related to the Jahn-Teller distortion should hybridize with spin
waves more strongly than people thought until now.
This is our main conclusion presented at the end of Sec.\ref{results}.

%%%%%%%%%%%%%%%%%%%%%%%%%%%%%%%%%%%%%%%%%%%%%%%%%%%%%%%%%%%%%%%%%%%%%%%%%%%%%%%%%%%%
\section{method}
\label{method}
We first explain the \qsgw\ method which is applied to 
calculations presented in this paper.

The $GW$ approximation (\GWA) is a perturbation method.
Generally speaking, we can perform \GWA\ from 
any one-body Hamiltonian $\H0$ including non-local static
potential $\veff(\bfr,\bfr')$ as
\begin{equation}
\H0 = \frac{-\nabla^2}{2m} + \veff(\bfr,\bfr').
\label{eq:defh0}
\end{equation}
The \GWA\ gives the self-energy $\Sigma(\bfr,\bfr',\omega)$
as a functional of $\H0$; the Hartree potential through the electron 
density is also given as a functional of $\H0$.
Thus \GWA\ defines a mapping from $\H0$ to $H(\omega)$, which is given
as $H(\omega) = \frac{-\nabla^2}{2m} + V^{\rm ext} +V^{\rm H}
+\Sigma(\bfr,\bfr',\omega)$. Here $V^{\rm ext}$ and $V^{\rm H}$ denote
the external potential from nucleus and the Hartree potential symbolically.
In other words, the \GWA\ gives a mapping 
from the non-interacting Green's
function $G_0=1/(\omega-\H0)$ to the interacting Green' function
$G=1/(\omega-H(\omega)$).

If we have a prescription to determine $\H0$ from $H(\omega)$, 
we can close a self-consistency cycle; that is,
$\H0 \to H(\omega) \to \H0 \to H(\omega) \to ...$
(or $G_0 \to G \to G_0 \to ...$, equivalently) can be
repeated until converged. One of the simplest example of the prescription is to use 
$H(\omega)$ at the Fermi energy $\efermi$, that is, $\H0=H(\efermi)$
for $H(\omega) \to \H0$. In practice, we take a better choice 
in the \qsgw\ method so as to remove the energy-dependence; we replace 
$\Sigma(\bfr,\bfr',\omega)$ with the static version of self-energy 
$\vxc(\bfr, \bfr')$, which is written as
\begin{eqnarray}
\vxc = \frac{1}{2}\sum_{ij} |\Psi_i\rangle 
  \left\{ {{\rm Re}[\Sigma(\ei)]_{ij}+{\rm Re}[\Sigma(\ej)]_{ij}} \right\}\langle\Psi_j|,
\label{eq:vxc}
\end{eqnarray}
where $\{\ei\}$ and $\{\Psi_i\}$ are eigenvalues and eigenfunction
of $H$, and $\Sigma(\ei)= \langle\Psi_i|\Sigma(\ei) |\Psi_j\rangle$.
Re[$X$] means taking only the Hermitian part of the matrix $X$.
With this $\vxc$, we can generate a new $\H0$,
that is, it gives a procedure $\H0 \to H \to \H0$. 
Thus we now have a self-consistency cycle. 
By construction, the eigenvalues of $\H0$ is in agreement with
the pole positions of $H(\omega)$. Thus the eigenvalue is directly
interpreted as the quasiparticle energies. This \qsgw\ method is
implemented as an extension of an all-electron
full-potential version of the $GW$ method \cite{kotani02} as detailed
in \cite{kotani07a}. 

Until now they have shown that \qsgw\ works well for kinds of
materials (see \cite{chantis07f,kotani07a} and references there in).
In \cite{kotani02}, Kotani and
van Schilfgaarde have shown that the ordinary one-shot $GW$ 
based on the LDA systematically gave too small band gaps 
even for semiconductors; this is
confirmed by other theorists \cite{shishkin07,bruneval06qsgw}.
Thus the self-consistency is essentially required to 
correct such too small band gaps\cite{mark06qsgw,kotani07a}. 
Furthermore, the adequacy of one-shot $GW$ is analyzed from kinds of view points in
\cite{mark06adeq}; e.g., it shows that the usual one-shot
$GW$ can not open the band gap for Ge as shown in its Fig.6 (band entanglement problem).
The self-consistency is especially important for such as 
transition metal oxides like \abmo\ when reliability of the LDA and the GGA is questionable.
We have shown that \qsgw\ works well for wide range of materials
including MnO and NiO \cite{Faleev04,mark06qsgw,kotani07a,kotani07b,
chantis06spin,chantis07f}. We observed still remaining discrepancies 
between the \qsgw\ and experimental band gaps, 
but they are systematic and may be mainly corrected by including
the electron-hole correlation in the screened coulomb interaction $W$
as shown by Shishkin, Marsman and Kresse \cite{shishkin07}. 

In this paper, we focus on these two objectives:
\begin{itemize}
\item[(i)] Difference of results in the \qsgw\ and in the LDA.
\item[(ii)] Are results in the \qsgw\ consistent with experiments?
      If not, what can the results mean?
\end{itemize}
For these objectives, we mainly treat the simplest cubic structure
of the perovskite, one formula unit per cell, for \abmo, 
where we set A as Ba in a virtual atom approximation, that is, 
La$_{1-x}$Ba$_x$ is treated as a virtual atom with the atomic 
number $Z=57-x$. 
We use $6 \times 6 \times 6$ ${\bf k}$ points in the 1st Brillouin 
zone (BZ) for integration. 
We also treat a rhombohedral case for $x=.3$
(two formula units per cell. Its structure is taken from 
\cite{dabrowski98}; angle of Mn-O-Mn is $\sim170$ degree)
to examine the effect due to the rotation
of oxygen-octahedra (this is not a Jahn-Teller distortion).
Neither phonon contributions nor the spin-orbit 
coupling are included in all presented calculations.

Because of the difficulty to apply the $GW$ method to systems with
localized $d$ electrons, even the one-shot $GW$ calculations were rarely
applied to \abmo\ until now. Within our knowledge, one is by Kino et al \cite{KinoLSMO03},
and the other is by Nohara et al \cite{nohara:064417}.
Both are only within the atomic sphere approximation for one-body potential.
In contrast, our method is in a full-potential method.
Thus our method here is superior to these works in this point,
and in the self-consistency in the \qsgw.

\newpage
%%% result %%%%%%%%%%%%%%%%%%%%%%%%%%%%%%%%%%%%%
\begin{widetext}
\section{results and discussions}
\label{results}
\begin{figure}[htbp]
\includegraphics[width=18cm]{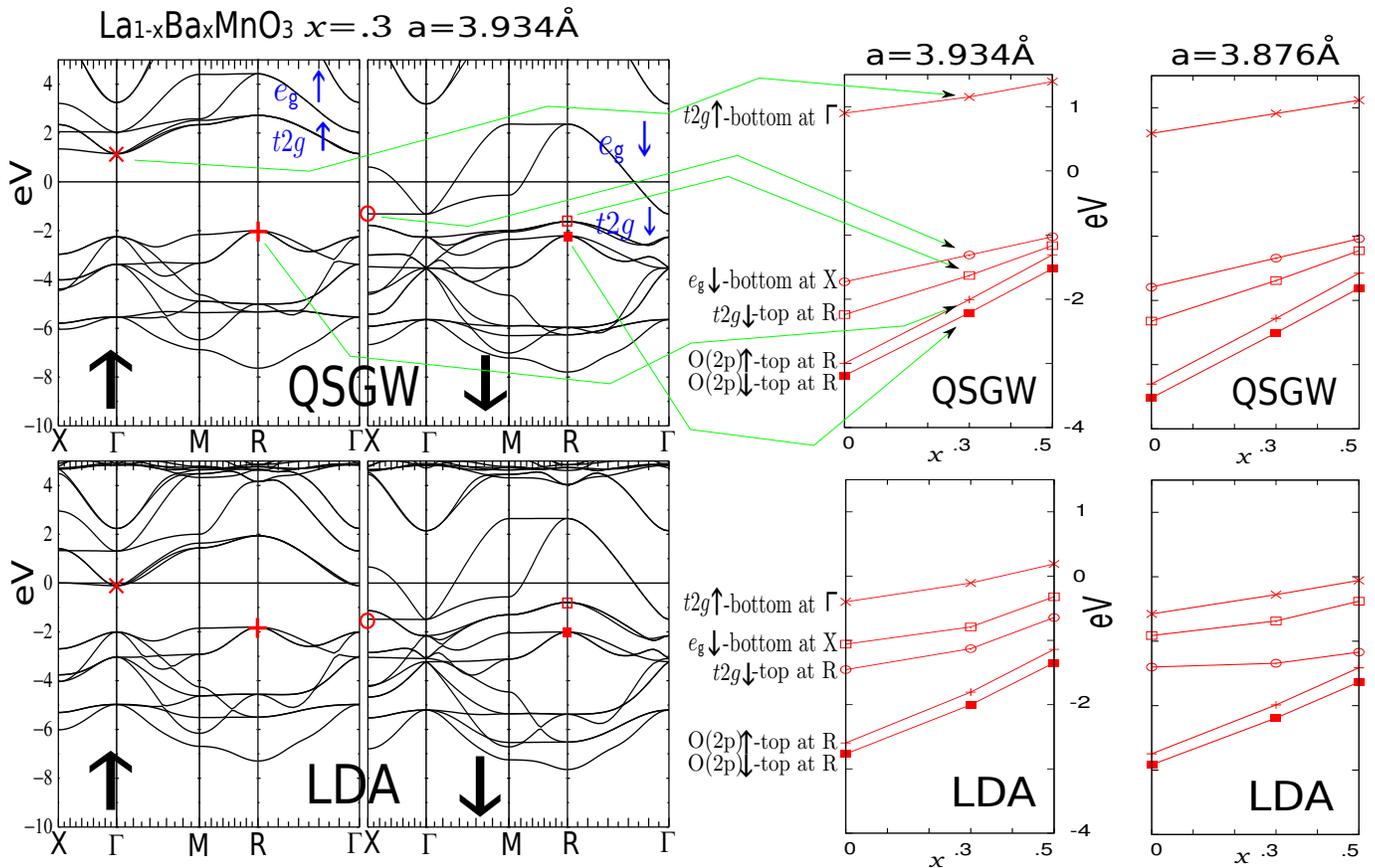}
\caption[]{(color online) The left panels are energy band at $x=0.3$ in \qsgw\ and in LDA
for the lattice constant, $a$=3.934\AA (as used in \cite{solovyev99})
in the black solid lines. The Fermi energy $\efermi$ is at 0 eV.
Right four panels show five typical eigenvalues 
for different $x$, not only for $a$=3.934\AA, but also for $a$=3.876\AA.
Those are shown by (red) symbols $\times$, $+$, $\bigcirc$, $\square$, and $\blacksquare$.
Correspondence to those in the left panels are indicated by thin (green) lines with arrows.}
\label{fig:band}
\end{figure}
\end{widetext}

In the left half of Fig.\ref{fig:band}, we compare energy bands in the \qsgw\ and in the LDA
at $x=0.3$ for lattice constant $a$=3.934\AA.
The energy bands are roughly assigned as O($2p$), \ttg, and \eg\ 
bands from the bottom. In its upper panels, we show labels \ttg\
and \eg\ to show the assignments.
The \qsgw\ gives a band gap in the minority spin ($\min$), that is, 
it is a half-metal, though the LDA does not.
This enhancement of half-metallic feature in \GWA\ is already reported even in the
one-shot $GW$ calculations by Kino et al \cite{KinoLSMO03}. Its
implication is emphasized in a recent review for half-metallic
ferromagnet by Katsnelson et al \cite{katsnelson:315}. 
The width of the \eg$\maj$ band in the \qsgw\ shows 
little difference from that in the LDA.
In the \qsgw, the \ttg$\maj$ band, which is hybridized with O($2p$)$\maj$, 
becomes narrower and deeper than that in the LDA. 

Right half of Fig.\ref{fig:band} shows five typical eigenvalues as
function of $x$, not only for $a$=3.934\AA, but also for $a$=3.876\AA.
In all cases treated here, \ttg$\min$-bottom at $\Gamma$ (bottom of
conduction band for $\min$) is above $\efermi$ ($\efermi$ is at 0 eV),
and O(2p)$\min$-top at R (top of valence band for $\min$) is 
below $\efermi$ in \qsgw. This means that it becomes fully-polarized half metals in
the \qsgw\ (thus the magnetic moment is given as $4-x \ \mu_{\rm B}$).
In contrast, the LDA gives a fully polarized half metal only when
$x=0.5$ for $a=3.934$\AA\ (\ttg$\min$-bottom is slightly above $\efermi$).
The eigenvalues of \ttg$\maj$-top at R are 
very close to that of O($2p$)$\maj$-top at R in the \qsgw, especially for large $x$.
Though the \qsgw\ eigenvalues show linear dependencies as a function of $x$,
the LDA does not. This is because the LDA has a small
occupancy for the \ttg$\min$ band.
%
%the eigenvalues denoted by $\bigcirc$, $+$, and $\blacksquare$
%(except $\times$ and $\square$ for \ttg) 
%in Fig.\ref{fig:band} in the LDA 
%agree well with those in the \qsgw;
%this is also true for their slope between $x=0.3$ and $x=0.5$. 
%At $a=3.876$\AA\ in the LDA, there are no gap in the 
%minority band for all $x$ ($\times$ in Fig.\ref{fig:band} is negative), 
%and the eigenvalue for \eg$\maj$ is only slightly dependent of $x$.

\begin{figure}[htbp]
\includegraphics[height=12cm,width=8cm]{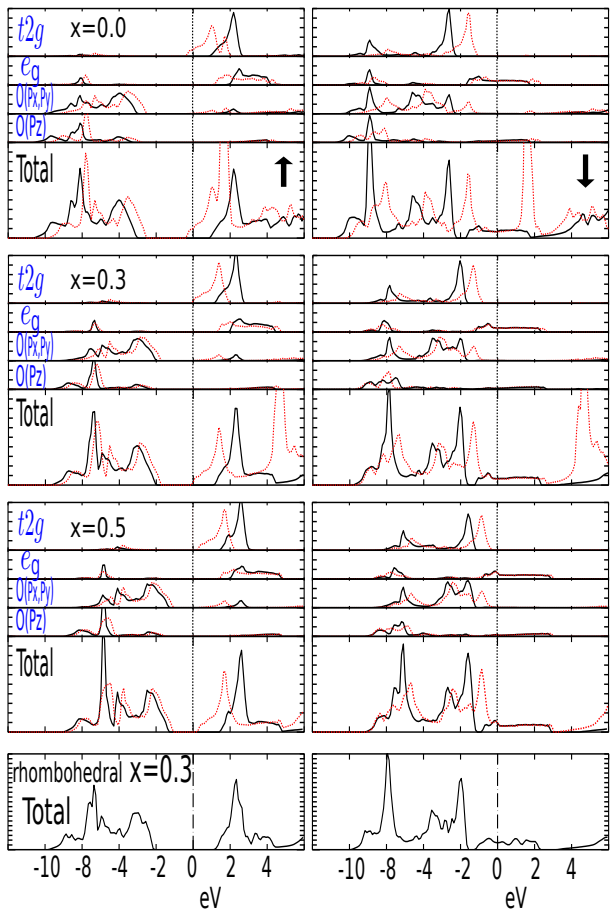}
\caption[]{(color online) Density of states in \qsgw\ (black solid line) 
and LDA(red dotted line) for $a$=3.934\AA. $\efermi$ is at 0 eV.
Left panels are for minority spin, and right panels are for majority spin. Four panels from the top to
the bottom are for $x=0.0$, $x=0.3$, $x=0.5$, and for the rhombohedral cases, respectively.
The $4f$ band in \qsgw\ is above the plotted region here in \qsgw.
O(p$_z$) denotes O($2p$) orbitals along Mn-O-Mn. O(p$_x$,p$_y$) are perpendicular to O(p$_z$).}
\label{fig:dos}
\end{figure}
Fig.~\ref{fig:dos} shows the corresponding total and partial density
of states (DOS). O(p$_z$) denote O($2p$) orbitals along Mn-O-Mn
bonding (for $\sigma$-bonding with \eg\ orbitals).
At first, La($4f$) level is located too low in LDA, 
at only $\sim$1.5eV above \efermi\ 
for $x=0$ \cite{sawada97}, though the \qsgw\ pushes 
it up to $\sim$\efermi $+ 10$ eV.
At $x=0$, all peak positions in \qsgw\ 
show some disagreement with those in the LDA.
This is due to the large difference in the occupation 
for the \ttg$\min$ band.
On the other hand, for $x=0.3, 0.5$, 
we see that differences of the total DOS are
mainly for a peak at $ \sim \efermi -2$eV in $\maj$,
and a peak at $\sim \efermi +2$eV in $\min$. 
The former difference is related to the the \ttg$\maj$ level.
If we push down the LDA \ttg$\maj$ level by $\sim$0.8eV, 
the occupied bands will be closer to those in \qsgw. 
The latter difference is related to both of the unoccupied Mn($3d$)
(\ttg$\min$ and \eg$\min$). The \qsgw\ pushes up
\ttg$\min$ and \eg$\min$ by $\sim$ 1eV, relative to the LDA results.
The experimental position of \ttg$\maj$ band 
is described well in the \qsgw\ than the LDA as follows.
Angle resolved photo emission spectroscopy (ARPES) by Liu et al \cite{liu00} 
concluded that Mn(3$d$) band (presumably due to \ttg$\maj$) 
is $\sim 1$eV deeper 
than the LDA result for La$_{.66}$Ca$_{0.33}$MnO$_3$. 
Chikamatsu et al \cite{chikamatsu06} also performed ARPES for La$_{0.6}$Sr$_{0.4}$MnO$_3$, 
showing that there is a flat dispersion around $\efermi  -2$eV.
These experiments for \ttg$\maj$ support the results in the \qsgw.
As for the positions of the unoccupied Mn(3$d$) bands,
no inverse photoemisson experiments are available to identify them 
though we give some discussion below when we show $\epsilon_2$.

As we see above,
the main difference between the \qsgw\ and the LDA
is interpreted as the difference of the exchange splitting 
for the \ttg\ band. Roughly speaking, center of \ttg$\maj$ and
\ttg$\min$ given in the LDA is kept in the \qsgw. 
Because of the larger exchange splitting, \qsgw\ shows large half-metallic band gaps.
In addition, the \eg$\min$ band is pushed up.
Based on the knowledge in other materials together with the above experiments,
we think that the \qsgw\ should give better description than the LDA.
Generally speaking, LDA can introduce two types of errors when we
identify the Kohn-Sham 
eigenvalues with the quasiparticle energies.
One is the $U$-type effect as in LDA+$U$. This is
onsite contribution for localized orbitals. 
The other is the underestimation of the band gap
for extended orbitals as in semiconductors 
(In the case of diatomic molecule, non-locality in the exchange term can distinguish bonding 
and anti-bonding orbitals, though onsite $U$ can not).
As seen in \cite{kotani07a}, \qsgw\ can correct 
these two at the same time without any parameters by hand.

\begin{figure}[htbp]
\includegraphics[width=8.5cm]{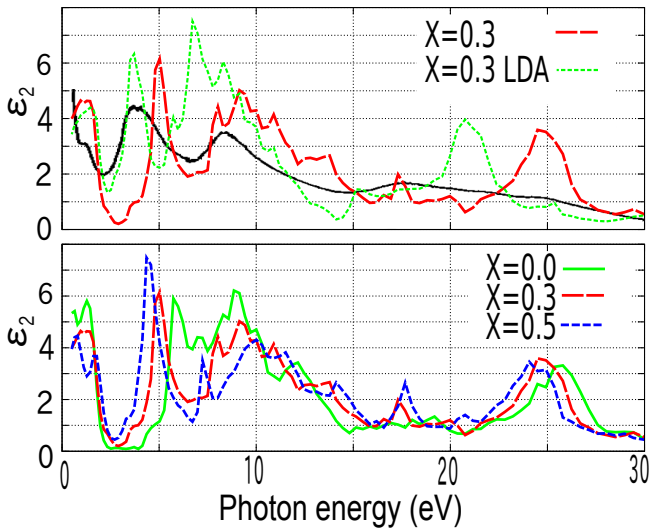}
\caption[]{(color online) Imaginary part of the dielectric function
$\epsilon_2$ for $a$=3.934\AA. Local-field correction is neglected but
it should be negligible as in the case of MnO and NiO \cite{kotani07a}.
Upper panel is to compare calculations in \qsgw\ and in LDA 
with an experiment
for La$_{0.3}$Sr$_{0.7}$MnO$_3$ \cite{okimoto97} by black solid line.
In lower panel, we show results in \qsgw\ for different $x$.
Because of a limit in our computational method,
$\epsilon_2$ for $<0.5$eV are not calculated
(In practice, our results are with the wave vector 
${\bf q}=\frac{2\pi}{a}(0,0,0.04)$ 
instead of ${\bf q}=0$, though we confirmed 
little changes even at ${\bf q}\to 0$.)}
\label{fig:eps}
\end{figure}
Fig.\ref{fig:eps} shows $\epsilon_2$ in comparison with the experiment
\cite{okimoto97} for La$_x$Sr$_{1-x}$MnO$_3$. 
The LDA seemingly gives reasonable agreement with the experiment.
For example, the peak position around 4eV,
which is mainly due to transitions within the $\min$ channel, 
seemingly give excellent agreement with the experiment (upper panel).
This is consistent with the conclusion by Ravindran et al \cite{ravindran02}. 
On the other hand, $\epsilon_2$ in the \qsgw\ 
makes peak positions located at higher energies 
than the experiment by $\sim$ 1eV.
However, this kind of disagreement is
what we observed in other materials\cite{kotani07a}, where 
we identified two causes making the difference;
(a) A little too high unoccupied quasiparticle 
energies in the \qsgw\, and 
(b) The excitonic effect which is related to the 
correlation motion of electrons and holes during the polarization
(we need to solve the Bethe-Salpeter equation).
As for (a), we have an empirical procedure to estimate the error due to
(a); a simple empirical linear mixing procedure 
of 80\% of $\vxc$(\req{eq:vxc}) with 20 \% of the 
LDA exchange-correlation practically worked well
as shown in \cite{chantis06spin,chantis07f}.
We have applied this to the case for $x=0.3$ and $a=3.934$.
Then the \ttg$\maj$ level ($\times$ in Fig.~\ref{fig:band})
is reduced from 1.15eV to 0.93eV.
This level of overestimation $0.2\sim0.3$eV is 
ordinary for band gaps of semiconductors \cite{mark06qsgw}. 
If this estimation is true, 
the main cause of the disagreement should be due to (b). 
We think this is likely because we expect large
excitonic effects due to localized electrons. 
At least, the disagreement in Fig.\ref{fig:eps} do not mean 
the inconsistency of the \qsgw\ results with the experiment, 
though we need further research on it in future.
In addition, the effect due to the virtual crystal approximation is unknown.
We have also calculated $\epsilon_2$ for rhombohedral structure, 
resulting in very small differences from that for the cubic one.
The lower panel in Fig.\ref{fig:eps} shows changes as a function of $x$.
Its tendency as function of $x$ (the first peak at
$\sim$5eV is shifted to lower energy,
and the magnitude of the second peak at $\sim$ 9eV is reduced 
for larger $x$) is consistent with the experiment \cite{okimoto97}.

%%%%%%%%%%%%%%%%%%%%%%%%%%%%%%%%%
\begin{figure}[htbp]
\includegraphics[width=8.5cm,height=7.5cm]{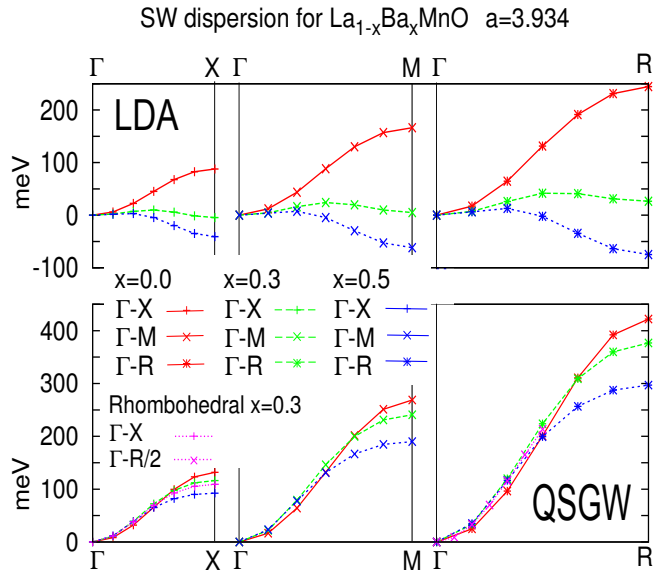}
\caption[]{(color online) Spin wave dispersion along
$\Gamma$-X, $\Gamma$-M, and $\Gamma$-R lines for $a$=3.934\AA. 
Negative energy means the unstable modes.
We also superpose the spin-wave dispersion in a rhombohedral case for
$x=0.3$ by pink lines (only $\Gamma$-X and $\Gamma$-R/2 in the \qsgw\ panels). 
It is almost on the cubic case, where R/2=(0.25,0.25,0.25) 
in cubic structure is on the BZ boundary of the rhombohedral structure.}
\label{fig:sw}
\end{figure}
Let us study the magnetic properties.
As discussed in \cite{solovyev99}, 
the exchange interaction
is mainly as the sum of the ferromagnetic contribution 
from the \eg\ bands, and the anti-ferromagnetic one from the \ttg\ bands. 
By the method in \cite{kotani08sw},
where the spin wave calculation based on the \qsgw\ 
reproduced experimental results very well for MnO and NiO,
we obtain the spin wave dispersions as shown in Fig.~\ref{fig:sw}.
The method is in a random-phase approximation to satisfy 
(spin wave energy) $\to 0$ at the wave vector $\bfq \to 0$.
In the LDA, the ferromagnetic ground state is stable at $x=0$, 
but it becomes unstable at $x\gtrsim 0.3$. This is 
consistent with the result by Soloveyv et al \cite{solovyev99},
though our LDA results are a little smaller than those for larger $x$.
On the other hand, we found that the ferromagnetic state is stable 
even at large $x$ in the \qsgw: Roughly speaking, the spin wave energies
in the \qsgw\ are about four times larger than experimental results \cite{ye07}.
We also show the spin waves for the rhombohedral case in Fig.~\ref{fig:sw}
(along $\Gamma$-X and along $\Gamma$-R/2),
but they are almost on the same line in the cubic case. 
This means that the rotation of the oxygen tetrahedra 
gives little effects for its magnetic properties. 
In order to check effects of overestimation of the exchange splitting 
in the \qsgw \cite{mark06qsgw,kotani07a,shishkin07},
we use the linear mixing of the $20$\% LDA exchange correlation
as we already explained when we discuss $\epsilon_2$, 
and calculate the spin-wave dispersion. 
Then it reduces the spin-wave dispersion by $\sim 11$ \%, 
thus our conclusion here is unchanged. 
Our results for the spin-wave dispersion in the \qsgw\ can be understood
as a result of the reduction of the anti-ferromagnetic contribution
of the \ttg\ bands because of their large exchange splitting.

As a summary, our result for the spin-wave dispersions in the \qsgw\ is clearly 
in contradiction to the experiments \cite{dai00,ye07,moussa07}.
In contrast, we have shown that the quasiparticle levels and
$\epsilon_2$ are reasonable and consistent with experiments.
Therefore we conjecture that it is necessary to include the
degree of freedom of phonons through the
magnon-phonon interaction so as to resolve the contradiction.
In fact, \cite{wood01,cheng08} had already suggested 
that the magnon-phonon interaction can change the spin-wave dispersions largely
by the strong hybridization with phonons of the Jahn-Teller distortion. 
In contrast, the magnon-phonon interaction was supposed to play a much smaller role for the
spin-wave dispersion in some experimental works than our result suggests:
For example, Dai et al \cite{dai00} and Ye et al 
\cite{ye07} claimed only the softening 
around the BZ boundaries are attributed to the hybridization:
Moussa claimed that the spin-wave dispersion is little affected 
by the magnon-phonon interaction \cite{moussa07}.

In conclusion, the \qsgw\ gives a very different picture 
from the LDA for the physics of \abmo.
The main difference from LDA is as for the magnitude 
of the exchange splitting for the \ttg\ band. 
It is $\sim$2 eV larger than that in the LDA.
\qsgw\ predicts a large gap in the minority spin (i.e., it is fully polarized).
Our results are consistent with the ARPES and the optical measurements,
but not with the spin-wave measurements. 
We think that this disagreement indicates a very strong hybridization
of spin wave with the Jahn-Teller type of phonons.
It should be necessary to evaluate its effect based on a reliable first-principle method.

We thank to Prof. M. van Schilfgaarde for helping us to 
use his code for full-potential Linear muffin-tin orbital method.
This work was supported by DOE contract DE-FG02-06ER46302, and by a Grant-in-Aid for Scientific
Research in Priority Areas "Development of New Quantum Simulators and Quantum Design"
(No.17064017) of The Ministry of Education, Culture, Sports, Science, and Technology, Japan.
We are also indebted to the Ira A. Fulton High Performance 
Computing Initiative. 
\bibliography{ecal}
\end{document}